# Microstructural origin of the simultaneous enhancements in strength and ductility of a nitrogen-doped high-entropy alloy

Xiaoxiang Wu[1*], Zhujun Sun[1], Wenqi Guo[2], Chang Liu[3], Yong-Qiang Yan[3], Yan-Ning Zhang[3], Yuji Ikeda[4,5], Fritz Körmann[4,5,6], Jörg Neugebauer[4], Zhiming Li[7], Baptiste Gault[4,8], Ge Wu[3*]

*1. School of Iron and Steel, Soochow University, 215137 Suzhou, China*

*2. Research Institute of Aero-Engine, Beihang University, 102206 Beijing, China*

*3. Center for Advancing Materials Performance from the Nanoscale (CAMP-Nano) and Hysitron Applied Research Center in China (HARCC), and Center for Alloy Innovation and Design (CAID), State Key Lab for Mechanical Behavior of Materials, Xi'an Jiaotong University, 710049 Xi'an, China*

*4. Max Planck Institute for Sustainable Materials, 40237 Düsseldorf, Germany*

*5. Institute for Materials Science, University of Stuttgart, 70569 Stuttgart, Germany*

*6. ICAMS, Ruhr-Universität Bochum, 44801 Bochum, Germany*

*7. State Key Laboratory of Powder Metallurgy, School of Materials Science and Engineering, Central South University, 410083 Changsha, China*

*8. Department of Materials, Royal School of Mine, Imperial College London, London, SW7 2AZ, U.K.*

Correspondence to: x.wu@suda.edu.cn (X. Wu); and gewuxjtu@xjtu.edu.cn (G. Wu)

**Abstract:**

As one of the most abundant interstitial elements, nitrogen (N) is effective in improving yield strength of metallic materials, due to interstitial solid solution strengthening. Doping N can substantially enhance the yield strength but often leads to a decreased ductility, revealing a strength-ductility trade-off phenomenon. Here, we simultaneously enhance the strength and ductility in a non-equiatomic CrMnFeCoNi high-entropy alloy via N alloying and unravel the underlying microscopic mechanisms. The N-doped alloy (1 at.% N) shows an excellent combination of higher yield strength (104% increase) and larger ductility (38% increase), with a two-stage strain hardening behavior, compared to the N-free alloy. Detailed transmission electron microscopy (TEM) analysis reveals that N-doping introduces short-range order (SRO) domains within the microstructure, leads to pronounced planar slip, and promotes the formation of nano-spaced (6-15 nm) stacking faults and deformation twins. Continuous generation and interaction of the fine-spaced SFs act as a strong barrier for dislocation movement and provide ample room for dislocation storage. The interaction of SRO with dislocations and the evolution of SFs ascribe to the first strain hardening stage, and the disordering of the SRO along with the activation of deformation twins are attributed to the second strain hardening stage. Our work shows that N-doping is effective in simultaneously improving the strength-ductility synergy and provides novel insights into alloy design with slightly elevating the SFE, and manipulating the ordered structure within the HEA.

# 1 Introduction

The simultaneous improvement of strength and ductility has long been sought after in the design of high-performance metallic materials [1–3]. Alloying with abundant elements and reducing the use of expensive and rare elements in metallic materials is an efficient way towards the sustainable development of novel metallic materials [4]. Nitrogen (N) is an earth-abundant element often added to metallic materials (e.g., steels and medium/high-entropy alloys (M/HEAs) to enhance mechanical performance. Compared to carbon (C), N is more effective in strengthening and holding, for some alloy types, advantages in low-temperature strengthening, impact toughness, and corrosion resistance [5–9]. Increasing the N content elevates yield strength through interstitial strengthening, yet this happens in some cases at the expense of ductility loss, especially when nitrides or intermetallic phases form [10,11]. More recently, doping N has been shown to simultaneously enhance the strength and ductility of FCC-based HEAs [10,12–15]. For example, Song et al. [12] introduced 1.8 at.% of N via additive manufacturing into heterogeneous structured FeCoNiCr alloys, overcoming the strength-ductility trade-off. He et al. [15] manipulated local chemical order in the partially crystallized Fe-Mn-Co-Cr-Ni alloy with heavy N-doping (3.2 at.%), reaching an ultrahigh yield strength of 1.34 GPa and a uniform elongation of 13.9%.

Yet, numerous questions remain, for example, how N affects the distribution of other solute elements and the evolution of crystalline defects, e.g., dislocations and stacking faults (SFs), the primary carrier of plastic deformation, to provide this improvement in properties. This is especially true for M/HEAs, which, compared to conventional alloys, present distinct features such as local compositional fluctuations [22,23], short-range order (SRO) [24–28], and distorted lattices [29,30] that lend them excellent mechanical properties. Interstitials are known to stimulate SRO formation and alter local SRO structure or local chemical fluctuation, such as ordered oxygen complex in oxygen-doped TiZrHfNb alloy [11], and Cr-N short-range order in the 1.75 at.% N-doped CoCrNi medium entropy alloy [7]. The change of microstructure (i.e., short/medium-range order, local chemical fluctuation or phase structure, etc.) can in turn modify the dislocation behavior and interact with other crystalline defects, such as SFs and deformation twins. Therefore, detailed investigations are required to understand how N-doping affects the movement and evolution of dislocations and other crystalline defects and thus the corresponding mechanical properties.

Decreasing the stacking fault energy (SFE) of the alloy is another effective way to obtain strength-ductility synergy [16–19], which is usually achieved by activating twinning-induced plasticity (TWIP) effect (with the SFE in the range of 20~40 mJ/m$^2$) and phase transformation-induced plasticity (TRIP) effect (when the SFE is lower than 20 mJ/m$^2$). Lu et al. [16] developed a novel high-N TWIP steel to simultaneously enhance the strength and ductility, thanks to the continuous decrease of SFE with high N amount and the ability to maintain the austenite stability. The impact of N on the SFE varies across different alloy systems. For instance, in the Fe-15Mn-2Cr-0.6C-xN TWIP alloy, N is found to increase the SFE linearly up to a concentration of 0.21 wt.% [20]. In contrast, in 316L austenitic stainless steel, the SFE decreases as the N concentration rises from 0.008 wt.% to 0.34 wt.%. [21]. This decrease in SFE is accompanied by a linear increase in strength and a reduction in elongation. The discrepancy of the effect of N on the SFE and mechanical properties raises two questions: first, whether N has the same controversial effect on the TRIP alloys, and second, whether achieving strength and ductility always requires a reduction of SFE in alloy design.

Here, we investigate the effect of N on a TRIP HEA and unravel the atomic-scale origins of how N enhances the strength-ductility synergy of an N-doped non-equiatomic CrMnFeCoNi HEA (($Cr_{20}Mn_{24}Fe_{30}Co_{20}Ni_6$)$_{0.99}$N$_{0.01}$). The 1 at.% N-doped alloy retains a single FCC phase. It shows an excellent combination of higher yield strength (104% increase) and larger ductility (38% increase) compared to its N-free reference alloy in the homogenized state. Combined with *ab initio* calculation and transmission electron microscopy (TEM) analysis, we reveal that the enhanced mechanical performance is achieved with an increase of SFE with 1 at.% of N-doping and a suppression of the TRIP effect. Detailed microscopic analysis at different length scales reveals that N shows no obvious portioning behavior in the N-doped HEA, but enhances the SRO structures that directly hinder dislocation motion and SF propagation during *in-situ* tensile deformation. A high density of fine-spaced SFs (6-15 nm), usually seen after severe plastic deformation [22] or cryogenic deformation [23,24], is associated with the enhanced strain hardening rate and corresponding extended ductility. We rationalize the effect of N-doping on the mechanical properties in terms of lattice distortion, change in SFE, and evolution of crystalline defects such as dislocations and SFs.

## 2. Materials and Methods

### 2.1 Alloy preparation.

The alloys studied in the present work have an atomic composition of $(Cr_{20}Mn_{24}Fe_{30}Co_{20}Ni_{6})_{0.99}N_{0.01}$ and $Cr_{20}Mn_{24}Fe_{30}Co_{20}Ni_{6}$, hereafter denoted as N-doped and N-free HEA, respectively. The alloys were prepared from pure elements in a vacuum induction furnace. N was added in the form of $Fe_2N$. Wet chemical analysis was utilized to confirm the concentration of the constituting elements after casting, and three different positions from cast ingots were measured to ensure accuracy. Table 1 shows the measured compositions after casting for the two alloys. These as-cast alloys were first hot-rolled at 950°C with a thickness reduction of 50%. Subsequent homogenization was conducted at 1200°C for 2 hours, followed by water quenching.

Table 1: Chemical composition for N-free and N-doped alloys (at.%)

|  | Fe | Mn | Ni | Cr | Co | N |
|---|---|---|---|---|---|---|
| N-free HEA | 30.4 | 23.8 | 5.9 | 19.9 | 19.8 | - |
| N-doped HEA | 30.6 | 23.2 | 5.9 | 19.7 | 19.5 | 1.1 |

### 2.2 Tensile testing.

The tensile testing was conducted on the dog-bone-shaped specimens using a Kammrath & Weiss tensile device. The tensile samples were prepared by electrical discharge machining, with a final thickness of 1 mm and a gauge area of 16 mm$^2$ (8 mm by length and 2 mm by width). Tensile testing was conducted at room temperature at a strain rate of $1.0 \times 10^{-3}$ s$^{-1}$. To precisely determine local strain evolution, the digital image correlation (DIC) method with the ARAMIS software (GOM GmbH) was utilized to track the strain evolution during the tensile process. Three specimens from each alloy were tested to ensure a reproducible tensile result, and the error bar for the tensile properties was raised from this statistic.

**2.3 Microstructural characterization**

The alloy phases after homogenization were first identified by X-ray diffraction (XRD), using X-ray equipment ISO-DEBYEFLEX 3003 with a cobalt source (CoK$\alpha$1, $\lambda$ =1.788 965Å) operated at 40 kV and 30 mA. The 2$\theta$ range for XRD measurement was 20-130° with a scan step of 0.03° and a counting time of 20 s/step. Bruker TOPAS software was further employed to analyze the XRD patterns, and Rietveld refinement was conducted to quantify the phase fractions.

A Zeiss-Sigma 500 VP was utilized for the backscattered electron (BSE) imaging and electron backscattered diffraction (EBSD) analysis, using a parameter of 15 kV and a step size of 80 nm. For a satisfactory EBSD analysis, samples were first ground using silicon carbide paper and followed by a 3-µm diamond suspension polishing. Further removal of the deformation layer introduced by mechanical grinding was realized by about 1-hour polishing with silica oxide particle suspension.

The microstructure after tensile deformation was investigated by TEM using a Jeol 2100 Plus equipped with a LaB$_6$ emission gun operated at 200 kV. High-resolution scanning transmission electron microscopy (HRSTEM) was carried out using a monochromated, probe-corrected ThermoFisher Titan 60-300 at a voltage of 300 kV. A semi-convergence angle of 17 mrad, probe current of 80 pA, and collection angles of the annular detector of 73 – 200 mrad were used for high-angle annular dark-field (HAADF) imaging. To reduce the influence of scan noise in HRSTEM HAADF, a serial acquisition scheme of 10-30 images with a dwell time of 2-5 µs. The *in-situ* TEM tension experiments at room temperature were conducted using a Gatan model 654 single-tilt straining holder with the normal e-beam density. *In-situ* studies were conducted utilizing JEM 2200 FS (from JEOL), operated at 200 kV. TEM foils were prepared by mechanical grinding to a thickness of around 100 $\mu$m, followed by electropolishing, using 5% perchloric acid in acetic acid at 17 °C. A site-specific plasma Xe+ focused ion beam (FIB) was used to prepare TEM lamellar samples with an orientation close to <011> for (S)TEM and *in-situ* TEM tension investigations. Final fine milling was conducted using 5 kV and 30 pA.

Specimens for APT measurement were prepared using a dual-beam FIB instrument (FEI Helios Nanolab 600i) and followed a routine atom probe tomography (APT) sample preparation procedure. The lift-out specimens were sharpened to a tip radius smaller than 80 nm with a low kV shower to remove Ga damage. The APT measurement was carried out in a LEAP 5000 XR from Cameca Instruments Inc.: at a high-voltage mode, with a pulse fraction of 15%, a pulse rate of 125 kHz, and a target evaporation rate of 1%. The specimens were kept at a base temperature of 60 K. The reconstruction of the 3D atom maps was analyzed using the CAMECA integrated visualization software IVAS 3.8.6.

**2.4 *Ab initio* calculations**

Both FCC and HCP phases of the $Cr_{20}Mn_{24}Fe_{30}Co_{20}Ni_6$ HEA were modeled using 54-atom supercells, with approximating the composition of the Cr:Mn:Fe:Co:Ni ratios with 11:13:16:11:3 as in Refs. [25–27]. The SFEs were computed using the first-order axial Ising model (AIM1) [28]. The ideal mixing of elements was modeled using special quasi-random structure (SQS) [29] configurations, which optimize the pair probabilities in the few nearest neighbor shells to approximate those in the ideal mixing state. To improve the computational performance, we constructed 76 configurations for each phase, and their energies were evaluated by taking the average over all the configurations. Solution energies of N atoms at octahedral interstitial sites in the alloy were evaluated by referencing the energy of an $N_2$ molecule, as done in Ref. [27]. To analyze the dependence on the local chemical environment around an N atom, we selected eight configurations among the 76 for each phase. We computed the N solution energies for all 54 octahedral sites for each of these eight configurations. In total, 864 sites were considered.

*Ab initio* density functional theory (DFT) calculations were performed using the plane-wave basis projector augmented wave (PAW) method [30] implemented in the VASP code [31–33]. The exchange-correlation energy was obtained within the generalized gradient approximation (GGA) of the Perdew–Burke–Ernzerhof (PBE) form [34]. The plane-wave cutoff energy was set to 400 eV. The reciprocal space was sampled by a Γ-centered 6 × 6 × 4 k-point mesh for the 54-atom models. The Methfessel–Paxton method [35] with a smearing width of 0.1 eV was applied. The 3d4s orbitals of Cr, Mn, Fe, Co, and Ni and the 2s2p orbitals of N were treated as valence states. Total energies were minimized until they converged within $1 \times 10^{-4}$ eV per simulation cell for each ionic step.

All calculations were performed considering collinear spin polarization. Based on the results previously found using the coherent potential approximation (CPA), all magnetic moments on Cr and Mn were initially set to be antiparallel to those on Fe, Co, and Ni [36]. During the self-consistent calculations, the local magnetic moments can adapt to their energetically preferred orientation, which is typically accompanied by spin flips on the Cr and Mn sites, as discussed in Ref [25].

The volumes and shapes of the supercells were kept fixed to the FCC lattice constant of 3.6 Å, close to the experimental value of CrMnFeCoNi [37–43]. The ideal *c/a* ratio of $(8/3)^{1/2} \approx 1.633$ was applied for the HCP phase, consistent with experimental findings [42,43]. HCP CrMnFeCoNi revealed *c/a* ratios close to the ideal one at ambient conditions. Metal atoms were initially put on the exact FCC and HCP lattice sites for the FCC and HCP supercell models, respectively. Ion relaxations were performed until the residual forces converged within $1 \times 10^{-2}$ eV/Å. An interstitial N atom was first placed at the geometric center, and then all the internal atomic positions were re-optimized.

## 3. Results

### 3.1 Initial microstructure prior to mechanical testing

We first examined the initial microstructure of the two alloys. Figure 1 presents the EBSD and XRD results for the two alloys prior to mechanical testing. Figures 1a-f show the inverse pole figure, the image quality map embedded with grain boundary distribution, and the Kernel Average Misorientation (KAM) map for N-free and N-doped HEAs, respectively. N-doping leads to similar grain size distributions (Figure 1g) compared to the N-free alloy (an average grain size of 46.3 μm for N-free HEA and 36.8 μm for the N-doped HEA, respectively) and does not result in the formation of precipitates or any second phase. Figure 1h compares XRD patterns of the N-free and N-doped HEAs. N-free HEA contains ~3 vol.% of the HCP phase, as indicated by the green tetragonal symbols. Both EBSD and XRD results suggest a single FCC structure for N-doped HEA. Adding 1 at.% of N suppresses HCP phase formation and stabilizes the FCC phase.

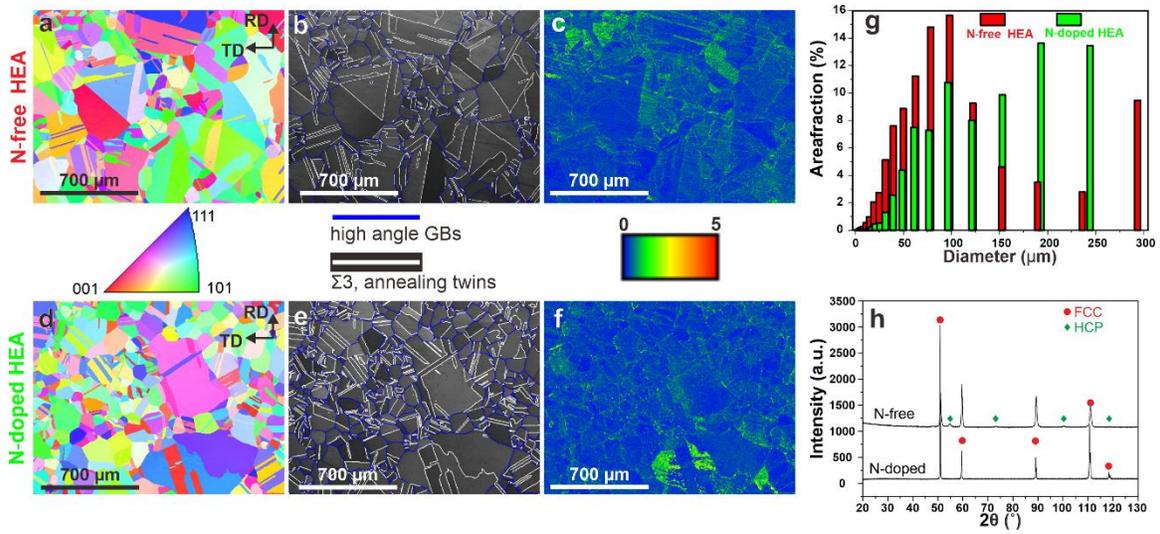

*Figure 1: EBSD and XRD analysis of the two alloys. N-free HEA: a. Inverse pole figure. b. Image quality figure overlayed with the high-angle distribution. c. KAM distribution. N-doped HEA: d. Inverse pole figure. e. Image quality figure overlaid with high-angle distribution. f. KAM distribution. g. Grain size distribution. h. XRD patterns.*

The initial microstructures of the N-free and N-doped HEAs were further investigated using TEM and HRSTEM. Figure 2a presents a representative grain of the N-free HEA close to [112] orientation, and the corresponding selected area diffraction pattern (Figure 2b) exhibits distinct extra diffraction spots (highlighted by dashed white circles and arrows) at 1/2{311} position along the [112] zone axis. These extra diffraction spots are typically regarded as evidence of short-range order (SRO) [44–46]. Figure 2c shows the corresponding HRSTEM HAADF image obtained from the region in Figure 2a, with its Fast Fourier Transform (FFT) patterns displayed in the lower right inset. The inverse FFT image in Figure 2d highlights the spatial distribution of the SRO regions. The initial microstructure of the N-doped HEA is shown in Figures 2 e-h. The diffraction pattern obtained from the region in Figure 2e also reveals similar extra diffraction spot, confirming the presence of SRO in the N-doped HEA. The representative inverse FFT image corresponding to the HRSTEM image in Figure 2g further illustrates the distribution of the SRO regions in Figure 2h. To quantitatively assess the influence of N addition on SRO, the intensity profiles (based on Figures 2b and 2f) and the spatial distribution of the SRO regions (based on Figures 2d and 2h) are compared in Figures 2i-j. The intensity analysis in Figure 2i clearly demonstrates that N-doping enhances the SRO intensity, while Figure 2j shows that the N-doped HEA contains a larger fraction of SRO regions. These microstructural observations provide direct evidence that the N addition significantly promotes the formation and degree of SRO in the alloy.

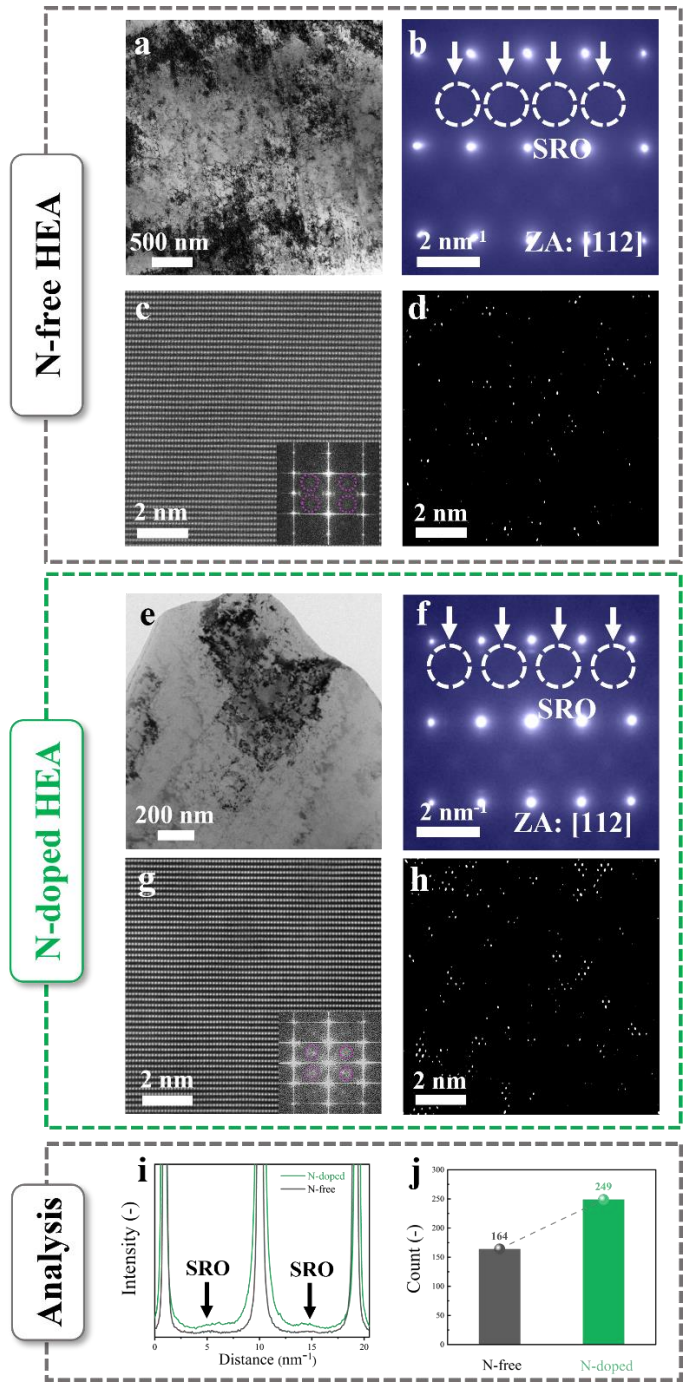

*Figure 2: TEM and HRSTEM analysis of the two alloys at the initial state. N-free HEA: a. TEM overview of the investigated region; b. Corresponding diffraction pattern of a; c. HRSTEM of the SRO region from a with the FFT inserted on the lower right; d. corresponding inverse FFT highlighting the SRO region; N-doped HEA: e. TEM overview of the investigated region; f. Corresponding diffraction pattern of e; g. HRSTEM of the SRO region from e with the FFT image inserted on the lower right; h. corresponding inverse FFT image highlighting the SRO region i: SRO intensity distribution analysis; j: SRO region comparison of the two alloys.*

### 3.2 Tensile properties

The engineering stress-strain curves for the two alloys are shown in Figure 3a, and the work hardening rate/true stress-true strain curves are displayed in Figure 3b, following Considère's criterion to establish the necking point and the uniform elongation. Here, with 1 at.% N doping, the yield strength increases by ~127.3% (from 165±3 MPa to 376±5 MPa), accompanied by a total elongation increase of ~38% (40±2% to 56±2%). The N-free HEA displays a conventional continuous decrease in the work hardening rate. In contrast, the N-doped HEA displays a two-stage strain hardening behavior. As revealed in Figure 3c, these significant improvements (by doping 1 at.% N) in mechanical properties outperform many other alloys with similar or higher interstitial contents, such as other FCC HEAs [10,12,47–52] or steels [53,54].

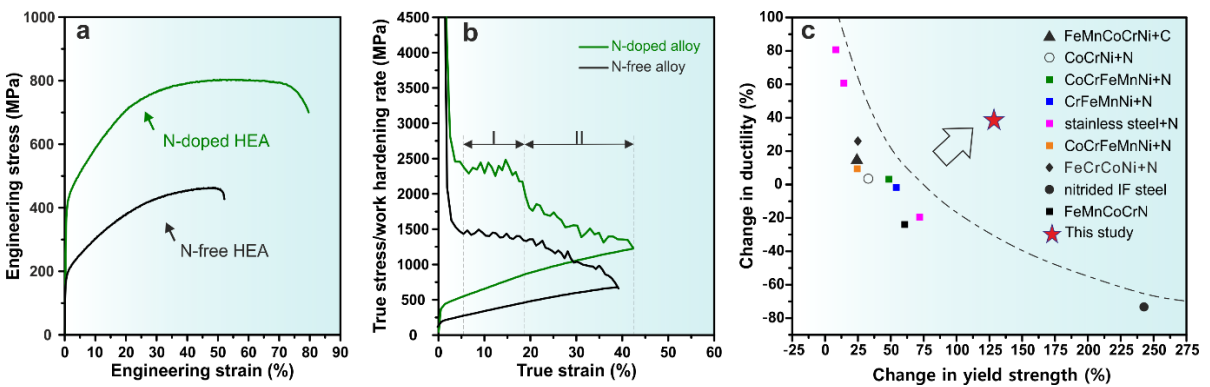

*Figure 3: Mechanical properties comparison. **a**. Engineering stress-strain curves of N-free and N-doped alloys. **b**. True stress-strain curve and work hardening rate-true strain curves of N-free and N-doped alloys. **c**. Change in yield strength and ductility of the current N-doped HEA as compared with other representative interstitial-doped alloys* [12,47–54].

## 3.3 Microstructural investigation after tensile testing

Microstructural investigation across different length scales was carried out to explore the mechanisms enabling the excellent enhancement of mechanical properties induced by N-doping. Post-mortem analysis of the N-free and N-doped HEAs at different tensile testing stages was carried out using TEM, STEM, and HRSTEM HAADF. Figure 4 compares the microstructural features of the two HEAs at an early stage of the tensile testing (a local strain of ~ 5%). Figure 4a shows one example of the distribution of random dislocations and planar traces (white dashed lines) in the N-free HEA. Figure 4b presents easy detection of long SFs (white triangles) and more frequent observation of planar slip traces (white dashed lines). In contrast, dislocation arrays in the N-doped HEA are distributed in a planar way under a {111} two-beam condition. Figure 4c shows an array of dislocations approaching the grain boundary, with a decreasing dislocation spacing close to the grain boundary region. Figure 4d displays another example of dislocation arrays at the lower right corner of the image, where paired dislocations can be clearly observed (green arrows). Both figures present dislocations clearly gliding in a planar way, and planar dislocations are often observed in alloys with low SFE or SRO [55].

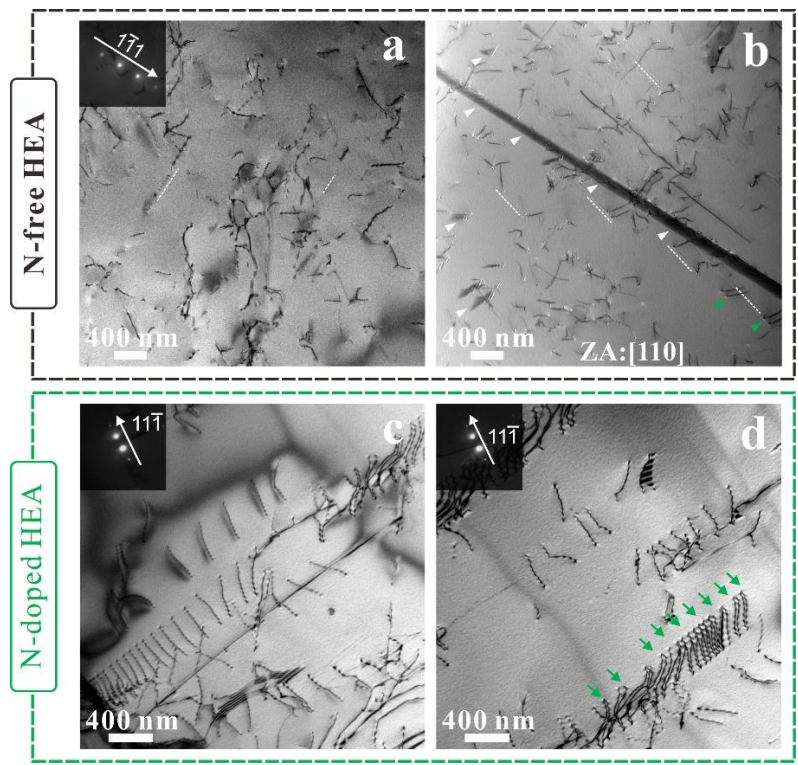

*Figure 4: TEM analysis of N-free and N-doped HEAs, showing the post-mortem microstructure at a local strain of ~5%. N-free HEA: a. Random dislocations and planar traces; b. Planar slip traces and SFs*. N-doped HEA: c. *Planar dislocations arrays approaching the grain boundary; d. Distribution of dislocation pairs (green arrows).*

With the local strain increasing to 50%, the microstructure is filled with a high density of dislocations and SFs. Figure 5a shows the dislocations saturated and forming slip bands in one direction. Another direction of slip bands and planar dislocations is present in Figure 5b. The slip bands are filled with short-segmented dislocations. Under multi-beam conditions with the beam parallel to the [011] zone, one variant of SF is viewed edge-on, as indicated by the orange arrows in Figure 6d, the enlarged view of the white rectangular region from Figure 5c. In-plane SFs can also be observed, as highlighted by the green arrows, intersecting with other SFs and dislocations. The high density of edge-on SFs, along with the slip bands, dynamically refines the grain and creates more interfaces, which hinder the movement of dislocations, as revealed in Figure 6. In addition, the confined dislocations in between further react and generate tangled dislocations, as highlighted with the green arrows in Figures 6a-b. The average spacing between the edge-on SFs in Figure 6a can be as fine as 10~15 nm. Figures 6c-d show the HRSTEM HAADF and BF images of the rectangular region from Figure 6a to reveal the atomic structure of the SF, where the SFs are intrinsic, lying on $(11\bar{1})$ plane.

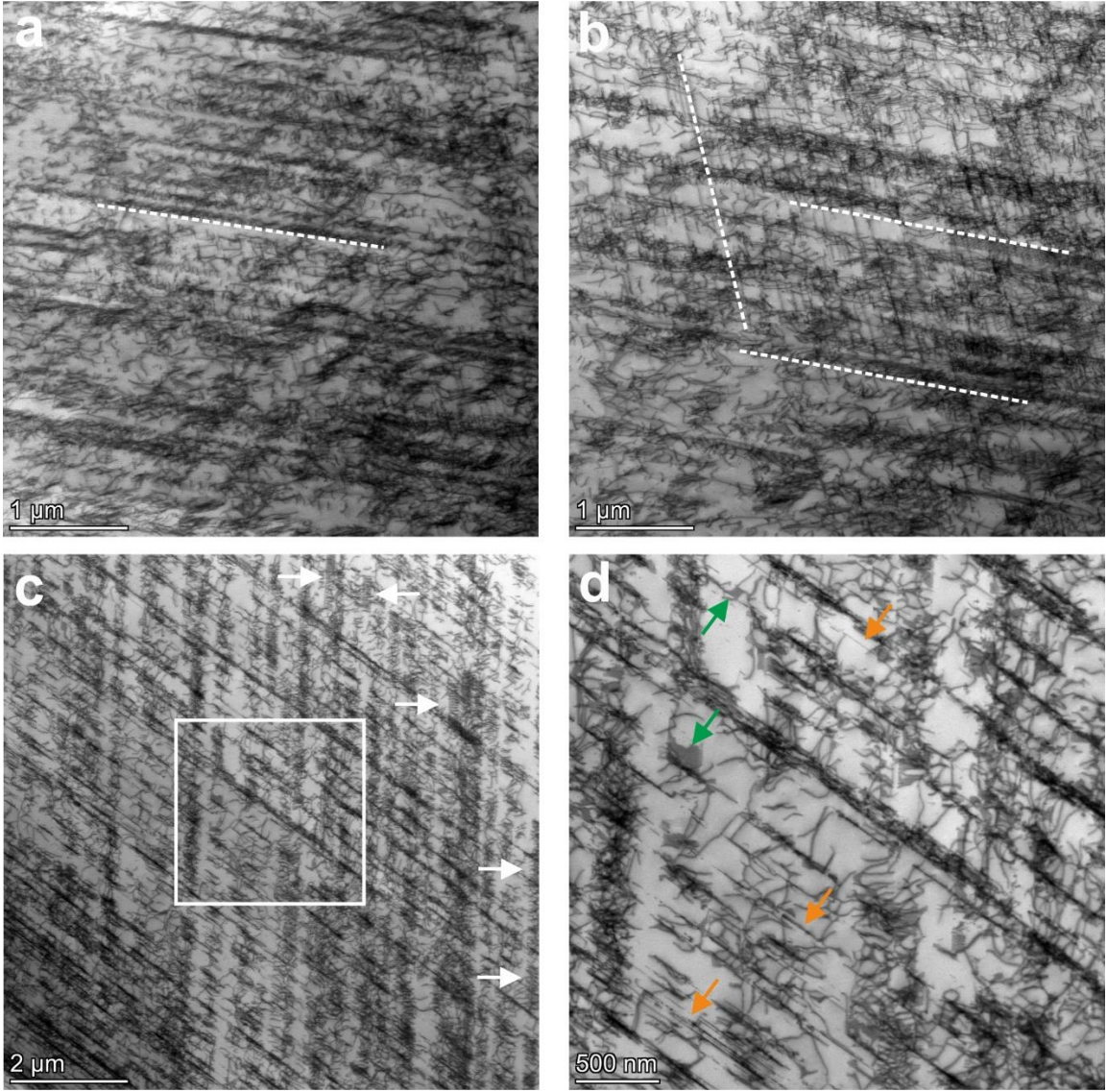

Figure 5: STEM analysis of the N-doped HEA, showing the post-mortem microstructure at a local strain of ~50%. **a**. Slip bands in one direction and high density of dislocations. **b**. Slip bands in two directions with planar dislocations in between. **c**. Edge-on SFs intersecting with slip bands. **d**. Enlarged view from Figure 5c highlighting the edge-on SF (orange arrow) and in-plane SF (green arrow).

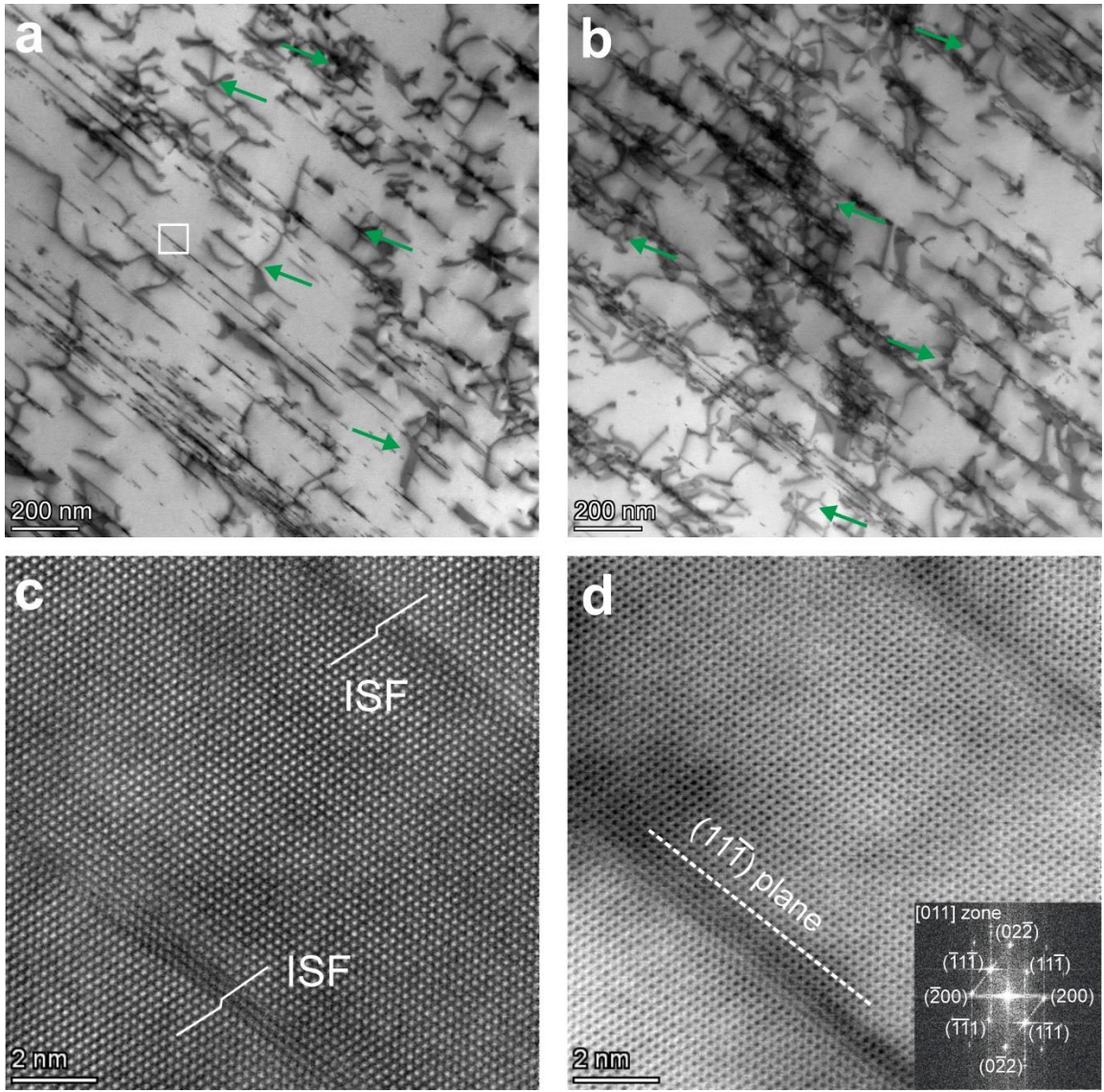

*Figure 6: STEM analysis at a local strain of 50% of the N-doped HEA. a. Edge-on and in-plane SFs interact with dislocations. b. Dislocations and SFs nodes structure. c. HRSTEM BF images with two intrinsic SFs (ISF). d. Corresponding HRSTEM HAADF image of Figure 6c.*

Upon further tensile deformation, SFs gradually evolve into deformation twin bundles with different thicknesses. The deformation twins are further confirmed with the extra symmetric diffraction spot. As shown in Figure 7, by using the extra spot (indicated by the yellow circle), dark field (DF) imaging highlights the distribution of deformation twins.

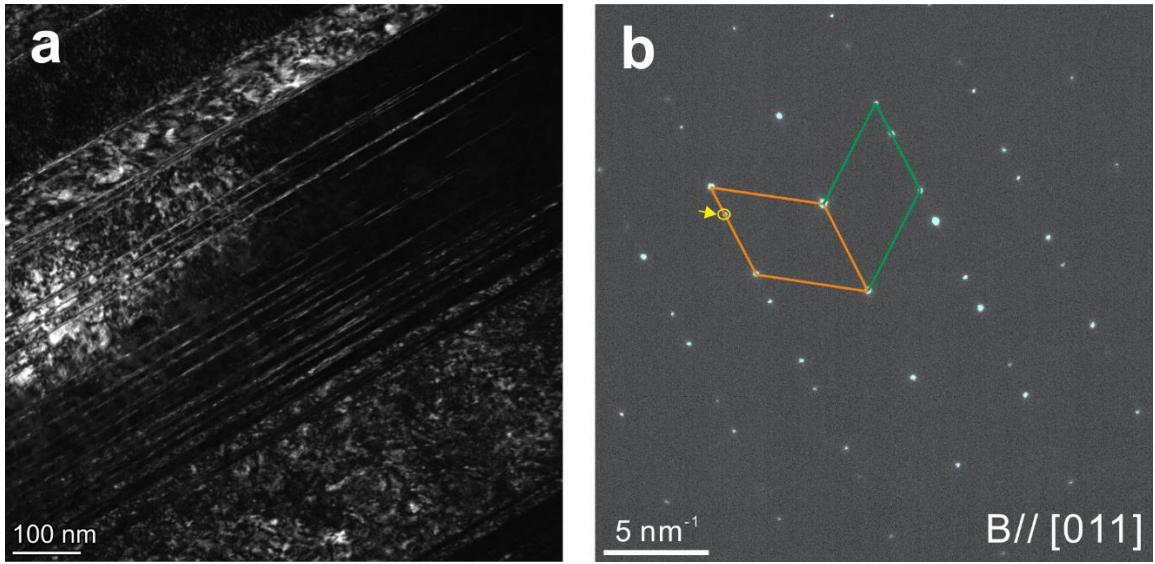

*Figure 7: TEM analysis of the N-doped HEA showing the deformation twins. a. TEM DF image showing the distribution of deformation twins. b. Corresponding selected area diffraction of the region from Figure 7a.*

In comparison, the N-free HEA presents a direct FCC to HCP phase transformation after tensile deformation. Figure 8 compares the dominant deformation mechanism of the N-doped and N-free HEAs by EBSD. Figure 8a shows the phase map of the N-free alloy after deformation. A significant fraction (71.5 vol.%) of FCC to HCP phase transformation was observed, and there is no obvious deformation twinning at this condition. This indicates the N-free HEA is more prone to triggering the TRIP effect. The KAM map in Figure 8c indicates a higher strain and geometrically necessary dislocation density at the interface between the FCC and HCP phases. In contrast, deformation twins prevailed in the N-doped HEA, as evidenced by the phase map in Figure 8d, which is consistent with the TEM analysis from Figure 7. In addition, the KAM value in the deformation twin regions is higher than in the TRIP region, evidencing that deformation twins can partition more strain and provide space for dislocation storage.

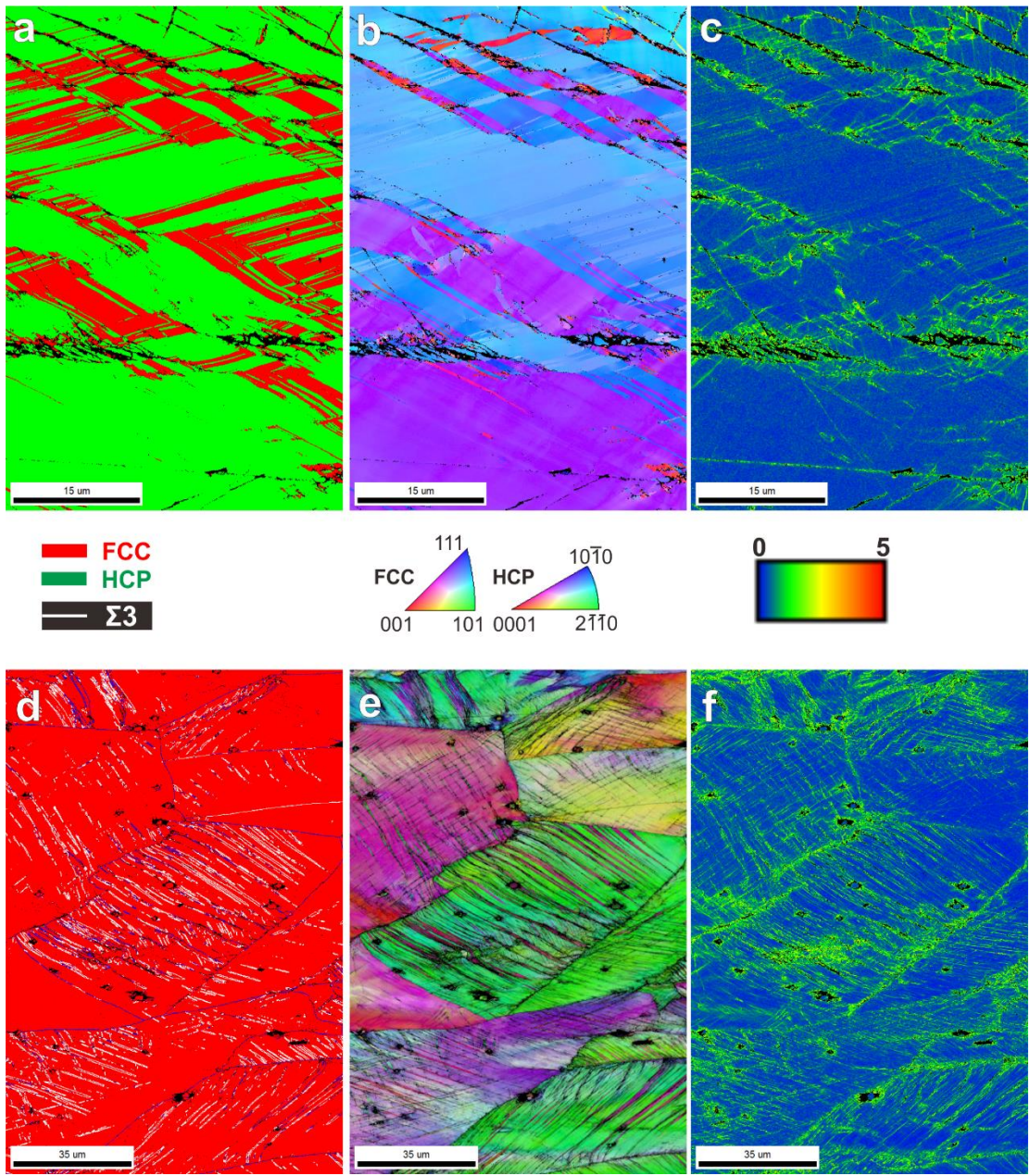

*Figure 8: Deformed microstructure by EBSD analysis. N-free HEA: **a**. phase map showing FCC phase (colored in red), HCP phase (colored in green), and deformation twin boundaries (colored in white); **b.** Inverse pole figure embedded with the image quality map; **c.** KAM map; N-doped HEA: **d**. phase map showing high-angle grain boundaries (colored in blue) and deformation twin boundaries (colored in white), no hcp phase is detected; **e.** inverse pole figure embedded with image quality map; **f.** KAM map.*

To unravel the role of N more specifically during the tensile process, we then performed *in-situ* tensile testing for the N-doped HEA. The as-prepared specimen for *in-situ* testing is shown in Supplementary Figure S1a with an orientation close to [011], and the region of interest is clean and free of defects. Upon loading, the region is quickly filled with a high density of dotted features on the scale of a few to tens of nanometers (see Supplementary Figure S1b). From the enlarged view of Supplementary Figures S1c and d, these features seem to be related to a change of strain, and some regions of the feature are related to SF. These strain-related dotted features and defects could be related to the SRO domains induced by N-doping. Upon loading, the external stress field leads to the changes of strain field around the SRO domains, promoting partial dislocation movement and facilitating the formation of the stacking faults. Since the *in-situ* testing was carried out in the TEM mode, the resolution and lack of EDS analysis limit the precise determination of the structural and compositional information of the SRO domains.

As the *in-situ* tensile testing proceeds, the SRO domains and generated SFs show an intriguing role, i.e., blocking the further movement of partial dislocations and SFs. Figure 8a shows the starting point of a moving SF during the *in-situ* tensile process. When the moving SF encounters one SRO domain, the propagation of the SF is hindered (see the snapshot from the *in-situ* TEM investigation in Figure 9b). Instead, a new parallel SF is activated, and the previously impeded SF would not move until sufficient stress is reached. The impeding process is present in the supplementary video 1. With the tensile process going on, a high density of fine and dense SFs is generated, and a second variant of SF is activated (see Figure 9c and supplementary video 2). The average SF spacing here is as fine as ~ 6 nm. These two variants of SFs continue to refine the microstructure, and their interaction leads to the formation of an SF network (Figure 10), which provides increased resistance to dislocation movement. The SF network eventually evolves into nano-deformation twins until fracture, as shown in Figures 9e-f and supplementary video 3.

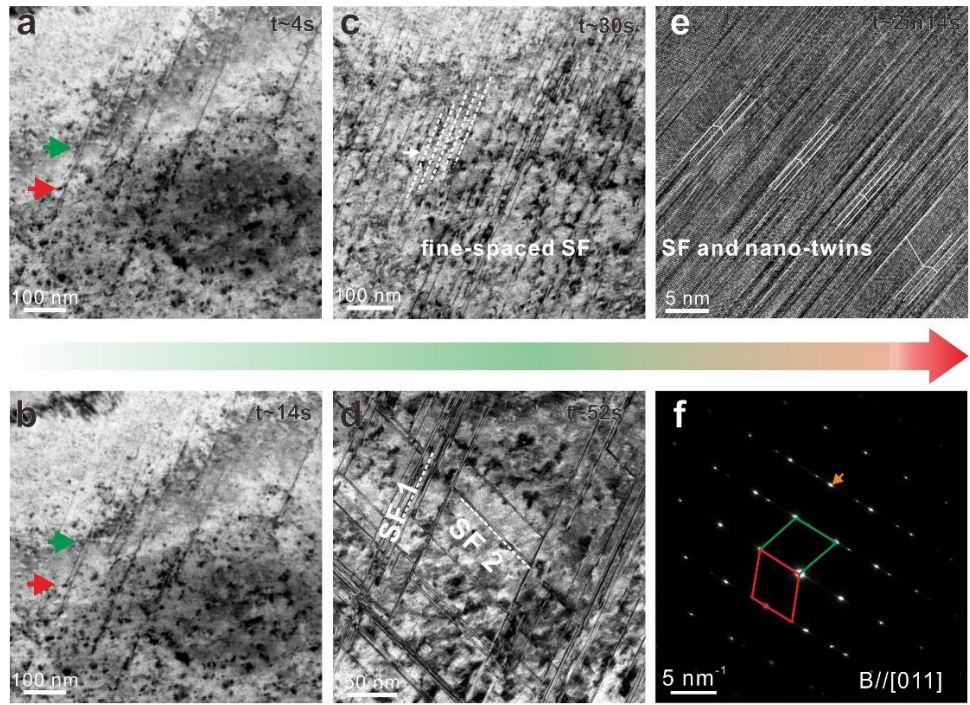

*Figure 9: In-situ TEM investigation. a. TEM snapshot showing the original position of the moving SF and the SRO domains. b. TEM snapshot showing the moving SF impeded by one SRO domain. c. Manipulation of the SFs with a high probability. d. Generation of second-variant SF, interacting with the first variant SF. e. HRTEM showing the evolution of SF and deformation twins. f. Corresponding diffraction pattern showing the twin symmetry spot and SF streak.*

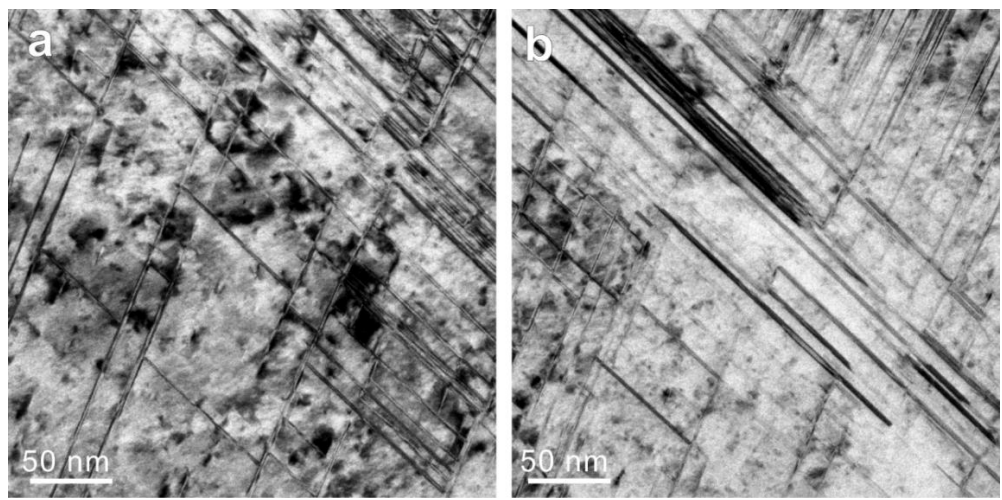

*Figure 10: In-situ TEM snapshots showing SF architectures. a. Two variants of SFs. b. Thickening of the SFs.*

## 4. Discussion

### 4.1 Strengthening and ductilizing via N-doping

In the present work, N-doped HEA achieves a simultaneous improvement of strength and ductility, compared to the N-free HEA, i.e., a yield strength increase of ~127.3% (from 165±3 MPa to 376±5 MPa) and a total elongation improvement of ~38% (402% to 562%). Meanwhile, the HEA is characterized by a salient two-stage strain hardening behavior, which is unique in the N-doped HEA. Based on the above detailed microscopic analysis, we first rationalize the excellent improvement in mechanical properties induced by 1 at.% of N-doping.

#### 4.1.1 Yield strength increment

The yield strength increase $\Delta\sigma_y$ is ~210 MPa per at.% of N, which is higher than compared of austenitic steels (140 MPa per at.%) [56] and hetero-structured N2.6 MEA (143.8 MPa) [10]. Since the present N-doped HEA contains SRO domains, their contribution to the yield strength has also been taken into account. The yield strength is normally ascribed to several contributing factors, such as lattice friction stress, solid solution, grain boundaries, etc., as shown in the following Equation 1 [57]:

$$\sigma_y = \sigma_{fr} + \sigma_{sro} + \sigma_{ss} + \sigma_{dis} + \sigma_{gb} + \sigma_{ppt}, \tag{1}$$

where $\sigma_{fr}$ is the lattice friction stress, and $\sigma_{ss}$, $\sigma_{dis}$, $\sigma_{gb}$, and $\sigma_{ppt}$ are the contributing stresses from solid solution, dislocations, grain boundaries, and precipitates. The substitutional solid solution strengthening term is commonly included in the friction stress term, since in the multicomponent environment of HEAs, it is difficult to define the solute and solvent atoms and interstitial solid solution strengthening can be folded into the lattice friction stress [57]. Here, 1 at.% of N doping yields a single FCC structure, excluding the precipitates' influence. The effect of the grain boundary is associated with the grain size distribution. However, according to the EBSD analysis of the initial microstructures in Figure 1, N-doping leads to slight grain refinement, with an average grain size distribution of 36.8 μm. The KAM distributions in Figure 1 show a relatively low and homogenous distribution of misorientations, which indicates no significant change in dislocation densities after N-doping in the initial state [58,59]. This rules out the effect of dislocation strengthening, and the yield strength increase Δσ$_y$ is mainly determined by Δσ$_{fr}$ and Δσ$_{SRO}$.

We then estimate the strengthening contribution of SRO. Following the model proposed by Kang et al. [60], the effect of SRO on the yield strength of the austenitic matrix can be estimated using the following Equation 2:

$$\Delta\sigma_{sro} = \Delta\sigma_y - 90 - 11.3 \times d^{-1/2} \tag{2}$$

Where d is the mean grain size in mm.

The yield strength Δσ$_y$ increase by N-doping is around 210 MPa, and the mean grain size of the N-doped HEA is ~36.8 µm. This calculation yields an increase in yield strength due to SRO of approximately 62 MPa. Given that the previously reported strengthening effect of nitrogen without SRO is ~140 MPa per at.% [10,56], it is thus reasonable to conclude that the additional strengthening observed in the present N-doped HEA is primarily attributable to the enhanced degree of SRO.

**4.1.2 Strain hardening mechanism of the N-doped HEA**

While the N-free HEA presents a continuous decrease in strain rate, the N-doped HEA shows a two-stage strain hardening. EBSD results in Figure 8 indicate that the N-free HEA is more metastable and characterized by an intensive displacive phase transformation from the FCC to the HCP phase. In contrast, the N-doped HEA shows higher FCC phase stability and possesses significant faulting behavior, as evidenced by the TEM analysis in Figures 5, 6, and 9. Based on the TEM and STEM investigations at different strain levels, the first strain hardening stage is attributed to the pinning effect of the SRO structures, accompanied by the evolution and propagation of planar slip bands as well as fine-spaced and high-density intersecting SFs. The SRO structures directly impede dislocation motion, as evidenced by the dislocation morphology in the early deformation stage (Figure 2b) and confirmed by the *in-situ* tensile testing. The resistance promotes further dislocation multiplication and accumulation, thereby contributing to the initial strain hardening of the alloy. Moreover, paired dislocations are observed within the slip bands (highlighted by the white arrows in Figure 5c), which are most likely associated with the presence of SRO structure, as SRO significantly influences the dislocation morphology [61]. The piled-up dislocations could in turn exert extra back stress to the source of initiating dislocations, and further enhancing strain hardening. In addition, the densely distributed SFs intersecting the planar slip bands impose extra resistance to dislocation motion, intensifying the overall hardening response of the alloy. In one aspect, SFs are analogous to twin boundaries regarding the role of refining the grains and prohibiting the motion of dislocations. The evolution of the parallel SFs continues to refine the grain and decrease the mean free path of mobile dislocations, triggering the dynamic Hall-Petch strengthening effect [62,63]. This aspect mainly refers to the parallel SFs, and an additional elastic repulsive force would be exerted upon the associated glissile dislocations with the SFs [64]. In another aspect, the fine-spaced SFs and the planar slip bands provide ample room for dislocation storage and dislocation/SFs interaction (e.g., Figures 5 and 6). This leads to efficient dislocation accumulation, contributing to the increase in ductility. Meanwhile, the interaction between dislocation and two variants of SFs leads to the formation of the SF node structure, which is similar to the Lomer-Cottrell lock (highlighted with green arrows in Figure 6). These dislocation-SF node structures provide enhanced resistance to dislocation movement and SF propagation, further hindering dislocation movement and thus strengthening the material.

Nevertheless, the obstructing effect of SRO is weaker than that of precipitates or grain boundaries. As deformation proceeds, the impeding role of SRO gradually diminishes, transitioning from an ordered to a disordered state [65]. In the meantime, with the increase of tensile stress, deformation twins are activated and set in to accommodate further tensile deformation. It is generally accepted that deformation twins can be formed by partial dislocation gliding on the successive {111} planes [66,67]. Partial dislocation gliding on {111} planes can lead to the formation of SFs, and the high density of SFs can act as embryos for deformation twins and promote deformation twining once the critical stress is reached [55]. The absence of SRO structure and the transition to deformation twinning result in the second stage strain hardening.

Compared to the TRIP effect in the N-free HEA, SF strengthening holds advantages in avoiding the formation hard yet less ductile HCP phase, mitigating the possibility of crack initiation and propagation. SF strengthening has also been observed in other alloy systems, such as CoCrNiW alloy with negative intrinsic SFE [64], Al$_{0.1}$CoCrFeNi alloy deformed at liquid nitrogen temperature [23], precipitate-hardened CrNiCo medium entropy alloy [70,71], and Cu-Al alloys with different grain sizes [72]. By tailoring the composition range and the thermal-processing ways, the morphology and stability of SFs can be manipulated, i.e., extremely fine SFs in multiple directions or long parallel SFs across the grains. Interestingly, the prevalence of SFs here is simply induced in the single-phase HEA with coarse grains after N-doping and tensile-tested at ambient temperatures. Since a high probability of SFs is normally associated with lowering SFE, we then explore the effect of N on SFE.

**4.2 Effect of N on the SFE**

The activation of different deformation mechanisms, i.e., dislocation slip, TRIP effect, and deformation twins, is closely related to SFE in various metallic materials. The suppression of the FCC to HCP phase transformation by deformation twins in the N-doped HEA indicates a change in the SFE after N doping. *Ab initio* calculations were performed to assess the influence of N on the SFEs by quantifying the N solution energies $\Delta E_{sol}$ in the FCC and HCP phases, based on the first-order axial Ising model (AIM1) [28]. Previous *ab initio* studies [27,73] found that interstitial C and N atoms in CrMnFeCoNi favor octahedral sites rather than tetrahedral sites. Therefore, we also considered the occupation of the octahedral sites by N. Figure 11 shows the distribution of the computed solution energies $\Delta E_{sol}$ of the interstitial N atoms in octahedral sites in the FCC and the HCP phases. The standard deviations of the $\Delta E_{sol}$ distributions are as large as 0.327 eV and 0.286 eV for the FCC and the HCP phases, respectively. This indicates that $\Delta E_{sol}$ strongly depends on the local chemical environment around the interstitial site. The average $\Delta E_{sol}$ is 0.132 eV lower in the FCC phase than in the HCP phase (Figure 11), and N partitioning in the FCC phase is energetically more favorable than in the HCP phase.

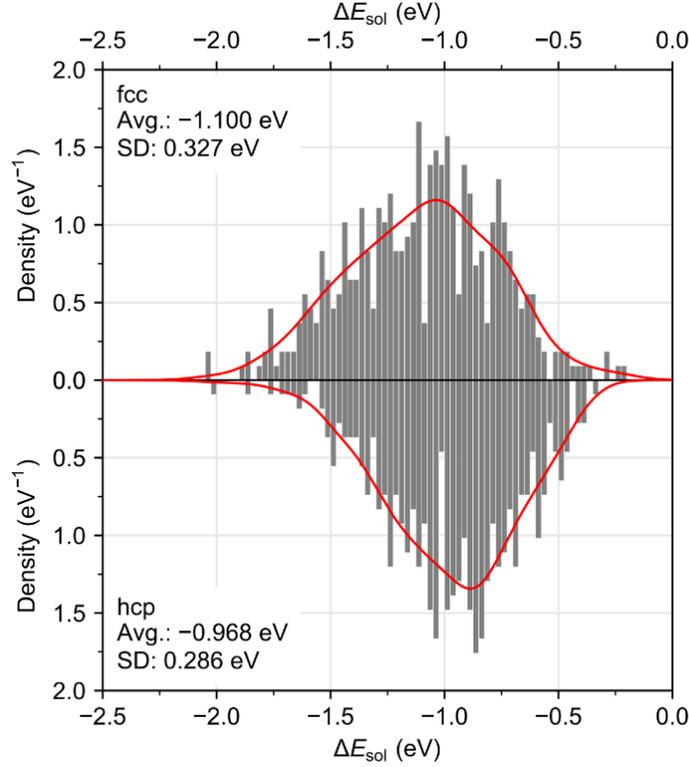

*Figure 11: Distribution of computed solution energies $\Delta E_{sol}$ of interstitial N atoms at the octahedral sites in $Cr_{20}Mn_{24}Fe_{30}Co_{20}Ni_6$ obtained from ab initio calculations. The upper and the lower panels show the results for the FCC and the HCP phases, respectively. The average (avg.) and the standard deviation (SD) of $\Delta E_{sol}$ are also shown in the panels. Red curves show the probability densities obtained by kernel density estimations.*

To derive the influence of N on the SFE, we further considered the thermodynamic limit where the N atoms occupy the interstitial sites according to the canonical ensemble for the $\Delta E_{sol}$ distributions for FCC and HCP [27,73]. Based on the AIM1, the SFE increases by 8 mJ/m$^2$ for 1 at.% of N atoms at 300 K, similar as found for CrMnFeCoNi [27] (8 mJ/m$^2$ for 1 at.% of N atoms). In addition, weak-beam dark-field TEM analysis was also used to measure the partial dislocation splitting and estimate the SFEs in the Supplementary Material. Due to the scattering of the splitting distance of the partial dislocations, the increase of SFE by N-doping measured from TEM (7.5±3 and 28.5±10 mJ/m$^2$ for N-free and N-doped HEA, respectively) is higher compared to the *ab initio* calculations.

Nevertheless, both *ab initio* calculations and the TEM measurements show an increased SFE after N-doping. This contradicts conventional wisdom with the idea of lowering the SFE to enhance the mechanical properties, as in N-doped TWIP alloys [16]. Here, the content of doped N is 1 at% (~0.281 wt.%) and is higher than the critical maximum content of 0.21 wt.% in stainless steels and TWIP alloys [16,20], and yet still increases the SFE of the TRIP HEA. As revealed by the previous microstructural evolutions, an increase in SFE results in the absence of TRIP effects but promotes a high density of fine-spaced SFs and deformation twins. This provides novel insight into high-performance alloy design, i.e., instead of driving for a continuous decrease of SFE, a slight enhancement of the SFE, accompanied by interstitial strengthening, can be more effective in improving the phase stability and strain hardening ability of the TRIP alloy.

Considering the intense planar slip and massive faulting events in the N-doped HEA, conventionally, planar slip is associated with three aspects, i.e., high frictional stress, low SFEs, and SRO effects [74,75]. With N-doping, the friction stress increases, induced by the interstitial strengthening, which impedes the dislocation movements. Yet, N-doping also leads to an increase in SFE, which rules out the effect of lowering SFE on planar slip of dislocations. This is consistent with the observed SRO structure in the initial structure.

**4.3 Comment on N-doping enhanced SRO and comparison with other interstitials**

The key role of N in the current work is to promote the formation of SRO structures, which in turn lead to the observed two-stage work-hardening behavior. Notably, such SRO structures have not been reported in carbon-doped or oxygen-doped FCC-based HEAs. In CrCoNi medium-entropy alloys, N doping has been shown to enhance strength while maintaining ductility [76], and no multi-stage strain hardening behavior has been observed. However, the strengthening mechanism has primarily been attributed to an increase in friction stress, without evidence of SRO effects. This discrepancy may further indicate the role of SRO during the tensile process and the tendency of SRO formation of the host HEAs. HEAs exhibit local chemical fluctuations in the solute distribution [77], and interstitial doping could further promote local compositional fluctuations or SRO structure [78]. This indicates that N can be utilized to tailor the degree of SRO in the FCC-structured HEAs. By contrast, other interstitials, such as carbon and boron, are less prone to the formation of the ordering structure; instead, they typically strengthen alloys through precipitation (e.g., carbides or borides) or by segregating to grain boundaries [79–81].

A similar SRO effect has been observed in boron-doped Cantor alloys deformed at cryogenic temperatures [82], but this SRO effect is absent at room temperature and promoted after tensile deformation specifically at 77K. In the BCC-structured TiZrHfNb HEAs, oxygen has been observed to form ordered complexes, a state between random distribution and precipitation [11, 80]. This kind of complex, however, is not present in the N-doped counterpart. It remains unclear how N concentration governs the extent of SRO formation, or whether changes in deformation temperature could amplify the N-induced SRO effect. These open questions call for further systematic investigations, which lie beyond the scope of the present study.

## 5. Summary and Conclusions

In summary, we present an efficient and cost-effective way to enhance the strength and ductility of quinary non-equiatomic HEAs simultaneously. Through detailed TEM and STEM analysis, the improvement of mechanical properties is associated with the formation of SRO structure and profound faulting behavior. We rationalize the mechanical behavior and deformation mechanisms of the N-doped HEA mainly in terms of frictional stress, formation of SRO domains, SRO to disordering transitions, and change in SFE. A better understanding of the effect of interstitials in HEAs offers valuable insight into the design of sustainable and high-performance alloys. The main summary and conclusions obtained from the current work are as follows:

(1) Doping 1 at.% of N into the non-equiatomic quinary HEA ($Cr_{20}Mn_{24}Fe_{30}Co_{20}Ni_6$) realizes simultaneously enhanced strength and ductility. The yield strength increases from 165±3 MPa to 376±5 MPa, accompanied by a 38 % improvement in ductility. N-doped HEA exhibits a strong two-stage strain-hardening behavior, and the enhancement of strength-ductility synergy compares well with other alloys.

(2) The strong strain-hardening behavior is closely related to the microstructure evolution. Detailed microscopic analysis shows that the deformation of the N-free HEA is dominated by the displacive phase transformation from FCC to HCP phase. The N-doped HEA is characterized by impeding behavior of SRO structures, planar slip and intensive and fine-spaced SFs hardening for the first strain hardening stage, where the continuous generation and interaction of the SFs provide a strong barrier for dislocation movement and ample room for dislocation accumulation. The second strain hardening stage is attributed to the disordering of the SRO structures and the transition from SF hardening to deformation twinning.

(3) As confirmed by *ab initio* calculation and TEM WBDF measurements, the enhanced mechanical performance induced by 1 at.% of N doping is associated with an increased SFE. Counter-intuitively, slightly elevating the SFE and suppressing the TRIP effect can act as efficient strengthening and ductilizing ways in the current TRIP alloy, thanks to the continuous generation of fine-spaced SFs and interaction between SF and deformation twins.

(4) *In-situ* TEM investigations reveal that N-doping-induced SROs are critical in hindering the movement of dislocations and SFs in the initial tensile stage. This observation also provides indirect evidence for chemical fluctuations in N-doped HEAs.

**Acknowledgment**

The authors would like to acknowledge the financial support from the National Natural Science Foundation of China (Grant Nos. 52101147, 52371162, 52361165617, 51971248, and 52271114), Suzhou Lab Open fund (Grant Nos. SZLAB-1108-2024-TS001), and Fundamental Research Funds for the Central Universities (xtr052025018). Y.I. and F.K. thank Andrei V. Ruban for providing the code to generate SQSs. D. R. and F.K. gratefully acknowledge support from the German Research Foundation (Deutsche Forschungsgemeinschaft, DFG) under the priority program 2006 "CCA-HEA." F.K. acknowledges the European Research Council (ERC) under the European Union's Horizon 2020 research and innovation program (grant agreement No. 865855). X. Wu would like to acknowledge the fruitful discussion with Dr. Y. Li, Dr. Y. L. Gong, Prof. D. Raabe from the Max-Planck Institute of Sustainable Materials, and Prof. Z. W. Wang from Central South University.

**Completing interests**

The authors declare no competing interests.